\newif\ifdoubleblind
\newif\iflackofspace
\lackofspacetrue    %

\newif\ifacmlackofspace
\newif\ifrelatedworktable

\ifdoubleblind
\documentclass[sigconf, screen, review, anonymous=true]{acmart}
\else
\documentclass[sigconf, screen]{acmart}
\fi

\usepackage[english]{babel}

\usepackage[utf8]{inputenc}
\usepackage{amsthm}

\usepackage[linesnumbered,ruled,vlined]{algorithm2e}

\SetKwComment{Comment}{\(\triangleright\)\ }{}   %
\SetKw{Return}{return}                          %

\usepackage{graphicx,subcaption} 
\usepackage{hyperref}
\usepackage[capitalise,noabbrev]{cleveref}
\usepackage{xcolor}
\usepackage{listings}   %
\usepackage{csquotes}   %
\usepackage{pifont} %
\newcommand{\cmark}{\ding{51}}%
\newcommand{\xmark}{\ding{55}}%

\newcommand*\xor{\veebar}    %
\usepackage{textcomp}

\iflackofspace
    \usepackage[ruled,vlined]{algorithm2e}
\else
    \usepackage[linesnumbered,ruled,vlined]{algorithm2e}
\fi
\SetAlCapNameFnt{\footnotesize}
\SetAlCapFnt{\footnotesize}

\usepackage{booktabs}
\usepackage{multirow}
\usepackage{tabularx}   %
\ifrelatedworktable
\usepackage{multirow}
\fi

\usepackage[font=footnotesize,labelfont=footnotesize]{caption}
\usepackage{enumitem}
\setlist[itemize]{noitemsep, topsep=0pt}
\usepackage{booktabs}
\let\originalbottomrule\bottomrule
\renewcommand{\bottomrule}{\addlinespace[0pt]\originalbottomrule}
\let\originalmidrule\midrule
\renewcommand{\midrule}{\addlinespace[0pt]\originalmidrule}
\newfloat{lstfloat}{htbp}{lop}
\floatname{lstfloat}{Listing}
\crefalias{lstfloat}{listing}

\usepackage{cprotect}
\usepackage{listings}
\usepackage{fancyvrb}

\usepackage{xcolor}

\definecolor{lightlightgray}{rgb}{0.9,0.9,0.9}
\definecolor{codegreen}{rgb}{0,0.6,0}
\definecolor{codegray}{rgb}{0.5,0.5,0.5}
\definecolor{codepurple}{rgb}{0.58,0,0.82}
\definecolor{codeblue}{rgb}{0.04,0.19,0.41}
\definecolor{codered}{rgb}{0.81,0.13,0.18}
\definecolor{backcolour}{rgb}{0.95,0.95,0.95}

\lstdefinestyle{cudastyle}{
backgroundcolor=\color{backcolour},   
commentstyle=\color{codegreen},
keywordstyle=\color{codered},
numberstyle=\tiny\color{codegray},
stringstyle=\color{codeblue}, %
language=C++,                   %
basicstyle=\scriptsize,       %
numbers=left,                 %
numberstyle=\scriptsize,      %
stepnumber=1,                 %
showspaces=false,             %
showstringspaces=false,       %
showtabs=false,               %
otherkeywords={__global__,__shared__,__constant__}, %
morekeywords={half,half2},
frame=none,                   %
framerule=0pt,
framesep=0pt, %
xleftmargin=0pt,
tabsize=2,                    %
captionpos=b,                 %
breaklines=true,              %
breakatwhitespace=true,      %
linewidth=1.03\linewidth,     %
}

\lstnewenvironment{cuda}[1][]{
  \vspace{-1mm}
  \lstset{style=cudastyle}
}{
  \vspace{-1mm}
}

\lstdefinestyle{pythonstyle}{
commentstyle=\color{codegreen},
keywordstyle=\color{codered},
numberstyle=\tiny\color{codegray},
stringstyle=\color{codeblue}, %
language=Python,              %
basicstyle=\scriptsize,       %
numbers=left,                 %
numberstyle=\scriptsize,      %
stepnumber=1,                 %
showspaces=false,             %
showstringspaces=false,       %
showtabs=false,               %
frame=none,                   %
frame=none,                   %
framerule=0pt,
framesep=0pt, %
xleftmargin=0pt,
breaklines=true,
tabsize=4,                    %
captionpos=b,                 %
escapeinside={\|}{\|},        %
linewidth=1.03\linewidth,       %
basicstyle=\fontsize{5.4}{2}\ttfamily, %
}%

\lstnewenvironment{python}[1][]{
  \vspace{-1mm}
  \lstset{style=pythonstyle}
}{
  \vspace{-1mm}
}

\lstdefinestyle{snippetstyle}{
backgroundcolor=\color{backcolour},   
commentstyle=\color{codegreen},
keywordstyle=\color{codered},
numberstyle=\tiny\color{codegray},
stringstyle=\color{codeblue}, %
language=C++,                   %
showspaces=false,             %
showstringspaces=false,       %
showtabs=false,               %
otherkeywords={__global__,__shared__,__constant__}, %
morekeywords={half,half2},
frame=none,                   %
framerule=0pt,
framesep=0pt, %
tabsize=4,                    %
captionpos=b,                 %
breaklines=true,              %
breakatwhitespace=false,      %
escapeinside={\|}{\|},        %
linewidth=\linewidth,     %
basicstyle=\footnotesize\ttfamily,
}

\AtBeginDocument{%
  }

\copyrightyear{2025}
\acmYear{2025}
\setcopyright{rightsretained}
\acmConference[ICPP '25]{54th International Conference on Parallel Processing}{September 08--11, 2025}{San Diego, CA, USA}
\acmBooktitle{54th International Conference on Parallel Processing (ICPP '25), September 08--11, 2025, San Diego, CA, USA}
\acmDOI{10.1145/3754598.3754601}
\acmISBN{979-8-4007-2074-1/25/09}

\begin{document}

\title{Efficient Construction of Large Search Spaces for Auto-Tuning}

\author{Floris-Jan Willemsen}
\email{f.q.willemsen@liacs.leidenuniv.nl}
\orcid{0000-0003-2295-8263}
\affiliation{%
    \institution{Leiden University}
    \city{Leiden}
    \country{the Netherlands}
}
\affiliation{%
    \institution{Netherlands eScience Center}
    \city{Amsterdam}
    \country{the Netherlands}
}

\author{Rob V. van Nieuwpoort}
\orcid{0000-0002-2947-9444}
\email{r.v.van.nieuwpoort@liacs.leidenuniv.nl}
\affiliation{%
  \institution{Leiden University}
  \city{Leiden}
  \country{the Netherlands}
}

\author{Ben van Werkhoven}
\orcid{0000-0002-7508-3272}
\email{b.van.werkhoven@liacs.leidenuniv.nl}
\affiliation{%
  \institution{Leiden University}
  \city{Leiden}
  \country{the Netherlands}
}
\affiliation{%
    \institution{Netherlands eScience Center}
    \city{Amsterdam}
    \country{the Netherlands}
}

\begin{abstract}

\iflackofspace
Automatic performance tuning, or auto-tuning, accelerates high-performance codes by exploring vast spaces of code variants. 
However, due to the large number of possible combinations and complex constraints, \emph{constructing} these search spaces can be a major bottleneck. 
Real-world applications have been encountered where the search space construction takes minutes to hours or even days. 
Current state-of-the-art techniques for search space construction, such as \emph{chain-of-trees}, lack a formal foundation and only perform adequately on a specific subset of search spaces. 

We show that search space construction for constraint-based auto-tuning can be reformulated as a \emph{Constraint Satisfaction Problem} (CSP). 
Building on this insight with a CSP solver, we develop a runtime parser that translates user-defined constraint functions into solver-optimal expressions, optimize the solver to exploit common structures in auto-tuning constraints, and integrate these and other advances in open-source tools. 
These contributions substantially improve performance and accessibility while preserving flexibility. 

We evaluate our approach using a diverse set of benchmarks, demonstrating that our optimized solver reduces construction time by four orders of magnitude versus brute-force enumeration, three orders of magnitude versus an unoptimized CSP solver, and one to two orders of magnitude versus leading auto-tuning frameworks built on chain-of-trees. 
We thus eliminate a critical scalability barrier for auto-tuning and provide a drop-in solution that enables the exploration of previously unattainable problem scales in auto-tuning and related domains.
\fi

\end{abstract}

\ifacmlackofspace
\else

\fi
\ifacmlackofspace
\else
\fi

\maketitle
\vspace{-0.1cm}
    \section{Introduction}
\label{sec:introduction}

Automatic performance tuning, or auto-tuning~\cite{balaprakash2017autotuning}, is a commonly applied technique in high-performance computing for optimizing programs towards a particular hardware architecture.
Auto-tuning allows developers to automate the process of exploring the myriad of implementation choices that arise in performance optimization, such as the number of threads, tile sizes used in loop blocking, and other code optimization parameters~\cite{lessonsLearnedGPU2020,hijma2023optimization}.
Many well-known examples of auto-tuned high-performance libraries and applications exist, such as FFTW~\cite{fftw1998} for Fast Fourier Transforms, or ATLAS~\cite{atlas2001} for linear algebra. 
At the heart of the auto-tuning method is a {\em search space} of functionally-equivalent {\em code variants} that is explored by an {\em optimization algorithm}.
These code variants can be generated by a compiler or using metaprogramming techniques, such as application-specific code generators or function templates.

Together, these code variants constitute vast search spaces that are infeasible to search by hand~\cite{pruning,scloccoAutoTuningDedispersionManyCore2014,CLTune} and would have to be searched over and over again as the application is executed on different hardware or different input data sets and sizes~\cite{CLBlast2018,vanWerkhoven2014optimizing,lawson2019cross,KTTBenchmark}.
Construction of the auto-tuning search space, used for enumerating and sampling different code variants, is a crucial factor in the performance of auto-tuning as search space sizes increase, and has therefore received a lot of attention recently~\cite{ATF,searchspaceATF,BaCO2024,pyATF}.

The difficulty in constructing search spaces for auto-tuning arises from the fact that not all code variants constitute valid implementations. 
In fact, many code variants that could potentially be generated violate so-called {\em constraints}. 
The constraints formulate dependencies between different tunable parameters in the code and often depend on limitations in both the program and the hardware. 
For example, when applying loop blocking, the tile size of the outer loop has to be a multiple of the tile size used in the inner loop, or a padding scheme in shared memory that only applies when shared memory is used and only for certain thread block dimensions.

In modern applications, where search spaces may contain millions or even billions of configurations, constructing the search space can take minutes to days, making search space construction a major bottleneck~\cite{BenchmarkingSuiteKerneltuners, heldensKernelLauncherLibrary2023, BaCO2024}. 
This is problematic when auto-tuning for a single target, but even more prohibitive on a diverse set of target input data and hardware, such as a BLAS library. 

To this end, Rasch et al.~\cite{ATF} have introduced a method, referred to as {\em chain-of-trees}, specifically designed for the purpose of efficiently constructing search spaces for constraint-based auto-tuning. 
The chain-of-trees approach starts with identifying groups of interdependent parameters. Two parameters are interdependent if they both occur in the syntax tree of the same constraint descriptor. %
For each parameter group, a tree is constructed that encodes all possible combinations of interdependent parameter values. %
Finally, the trees are linked together to form a chain-of-trees.
The chain-of-trees method is widely considered to be state-of-the-art and has been integrated into several auto-tuning frameworks, including ATF~\cite{searchspaceATF}, BaCO~\cite{BaCO2024}, PyATF~\cite{pyATF}, and KTT~\cite{KTTSoftwareX}.

In this paper, instead of adopting a customized solution for constraint-based auto-tuning, such as chain-of-trees, we investigate the use of methods with a more robust mathematical foundation. 
In particular, we show that the problem of search space construction in constraint-based auto-tuning can be automatically reduced to a Constraint Satisfaction Problem (CSP). 
To enable the use of CSP in a state-of-the-art auto-tuner, we employ various techniques, including run-time compilation to translate user-defined constraint functions to representations that can be used directly in existing CSP solvers, as well as major improvements to the performance of such solvers.
We evaluate the CSP and chain-of-trees methods on a wide variety of search spaces, including the search spaces used by Rasch et al.~\cite{searchspaceATF}.
Evaluation shows that our optimized CSP-based search space construction method is four orders of magnitude faster than brute force construction, three orders of magnitude faster than an unoptimized CSP solver, and one to two orders of magnitude faster than other state-of-the-art auto-tuning frameworks. 

Our work has been integrated into the auto-tuning framework Kernel Tuner~\cite{vanwerkhovenKernelTunerSearchoptimizing2019} and a Python-based CSP solver named python-constraint, both existing open-source projects with a substantial number of users, to benefit both communities. %

The rest of this paper is structured as follows.
\Cref{sec:background} provides a high-level introduction to the role of constraints in auto-tuning.
\Cref{sec:related_work} discusses related work.
\Cref{sec:searchspace_construction} describes the design and implementation of our method.
In \cref{sec:evaluation}, we evaluate the efficiency and scalability of our optimized CSP-based approach against various state-of-the-art solutions, and 
\cref{sec:conclusion_futurework} concludes.

    \section{Constraint-based Auto-tuning}\label{sec:background}

This section provides a general introduction to auto-tuning using an example kernel to illustrate how constraints arise when creating tunable applications for modern highly parallel architectures.

We will use the Hotspot kernel from the BAT~\cite{BenchmarkingSuiteKerneltuners} benchmark suite of tunable kernels as an example. 
This Hotspot kernel, adapted from the Rodinia Benchmark suite~\cite{cheRodiniaBenchmarkSuite2009}, simulates heat dissipation in a microprocessor based on the processor's architectural floor plan, thermal resistance, ambient temperature, and simulated power currents. 
\Cref{code:hotspot-naive} shows the Hotspot kernel code in HIP/CUDA, simplified for readability by removing bound checks.

\begin{lstfloat}[tb]
\begin{cuda}
__global__ void calculate_temp(float **Tout, float **Tin, float Tambient, float **Power, float3 R, float dt) {
  int x = blockIdx.x * blockDim.x + threadIdx.x;
  int y = blockIdx.y * blockDim.y + threadIdx.y;

  Tout[y][x] = Tin[y][x] + dt * ( power[y][x] + 
        (Tin[y+1][x  ] + Tin[y-1][x  ] - 2.0*Tin[y][x]) * R.y + 
        (Tin[y  ][x+1] + Tin[y  ][x+1] - 2.0*Tin[y][x]) * R.x +
        (Tambient - Tin[y][x]) * R.z); 
}
\end{cuda}
\caption{Example Hotspot kernel in HIP/CUDA.}
\vspace{-0.5cm}
\label{code:hotspot-naive}
\end{lstfloat}

The kernel uses a two-dimensional thread block to calculate the temperature of the chip at the new time step, where each
thread computes one value in the output matrix \texttt{Tout}. In this kernel, we can change the thread block x and y dimensions
without affecting the output, as long as we create enough thread blocks to cover the entire problem domain.
In other words, the thread block dimensions in x and y are {\em tunable parameters} that affect the performance but not the 
outcome. %
The optimal values for these parameters are highly specific to the kernel, the target hardware platform, and the input problem.
We thus select a wide range of values for both parameters.

However, to ensure sufficient parallelism, we need to make sure that the thread block contains at least 32 threads. 
In addition, most parallel architectures pose an upper bound on the number of threads per block, which can be queried before tuning. 
For simplicity, we here assume that this limit is 1024 threads.
These restrictions on the two parameters together form a constraint, namely \\ \texttt{32 <= thread\_block\_x * thread\_block\_y <= 1024}.

Different autotuners support different formats for specifying constraints, as illustrated in \cref{code:constraints}. 
ATF and PyATF both define constraints directly on the last parameter of a group of interdependent parameters, whereas KTT and Kernel Tuner define constraints separately from the tunable parameters. 
Kernel Tuner allows constraints to be defined as lambda functions or using string expressions. 
The model that all tuners follow is that the lambda functions, or string expressions, should evaluate to \texttt{True} for any specific combination of tunable parameter values to be considered a {\em valid} candidate solution, or code variant, in the search space.

\begin{lstfloat}[tb]
\begin{cuda}
// KTT
auto minWGConstraint = [](const std::vector<uint64_t>& v) {return v[0] * v[1] >= 32;};
tuner.AddConstraint(kernel, {"thread_block_x", "thread_block_y"}, minWGConstraint);
auto maxWGConstraint = [](const std::vector<uint64_t>& v) {return v[0] * v[1] <= 1024;};
tuner.AddConstraint(kernel, {"thread_block_x", "thread_block_y"}, maxWGConstraint);

// ATF combines tunable parameter and constraint declaration
auto thread_block_y = atf::tuning_parameter("thread_block_y", atf::interval<size_t>(1, N), [&](size_t thread_block_x){ return (thread_block_x*thread_block_y >= 32 &&     thread_block_x*thread_block_y <= 1024) });

// PyATF uses an interface similar to ATF, but in Python
thread_block_y = TP('thread_block_y', Interval(1, M), lambda thread_block_y, thread_block_x: thread_block_x*thread_block_y >= 32 and thread_block_x*thread_block_y <= 1024)

// Kernel Tuner (lambda-based constraint API)
constraint = lambda p: 32 <= p["thread_block_x"] * p["thread_block_y"] <= 1024

// Kernel Tuner (string-based constraint API)
constraint = "32 <= block_size_x*block_size_y <= 1024"
\end{cuda}
\vspace{-0.4cm}
\caption{Example of constraints specification in different tuners.}
\label{code:constraints}
\end{lstfloat}

The constraint defined in Listing~\ref{code:constraints} only involves the thread block dimensions. 
However, the fully optimized version of the Hotspot kernel contains many more tunable parameters and constraints. 
For example, the number of output elements computed by each thread can be varied in both the x- and y-dimensions, introducing two more tunable parameters. 
Another tunable parameter (\verb|sh_power|) controls whether or not to cache the \texttt{Power} values in {\em shared memory}. 
Together, these parameters form another constraint, namely \texttt{(thread\_block\_x * work\_per\_thread\_x) * (thread\_block\_y * work\_per\_thread\_y) * sh\_power * 4 <= max\_shared\_memory\_per\_block} to avoid exceeding the maximum amount of shared memory allowed per block in bytes. %
The full Hotspot kernel is even more complex and also implements temporal tiling, partial loop unrolling, and double buffering of the temperature field in shared memory, resulting in several more complex constraints. 
\iflackofspace
For further specification of the kernel see~\cite{BenchmarkingSuiteKerneltuners}.
\fi

    \section{Related Work}\label{sec:related_work}

Table~\ref{tab:related_work} shows an overview of support for specifying constraints, as well as the method used for search space construction in related auto-tuning frameworks. 
As we can see, some frameworks rely on brute force search space construction, \textit{ytopt} and \textit{GPTune} use \verb|ConfigSpace| and \verb|scikit-optimize.space| respectively, while the chain-of-trees method is most commonly used. 
We will briefly discuss the pros and cons of these different approaches.

\begin{table}[tbp]
\caption{Overview of constraint support and search space construction methods in related work and this work (Kernel Tuner). 
$\star$ While ytopt and GPTune are actively maintained, dependencies \textit{ConfigSpace} and \textit{scikit-optimize} are not.}
\label{tab:related_work}
\scriptsize
\vspace{-0.3cm}
\begin{tabular}{l|p{1cm}|p{0.9cm}|p{1.5cm}|p{1.8cm}}
\toprule
Tuner	& Open Source	& Actively developed	& Constraints API	& Search Space Construction	\\
\midrule
AUMA~\cite{AUMA}	& \cmark	& \xmark	& n/a	& external	\\
CLTune~\cite{CLTune}	& \cmark	& \xmark	& C++	& brute-force	\\
OpenTuner~\cite{OpenTuner}	& \cmark	& \xmark	& n/a	& brute-force	\\
ytopt~\cite{ytopt}	& \cmark	& \cmark$\star$	& Python	& ConfigSpace	\\
GPTune~\cite{liuGPTuneMultitaskLearning2021}	& \cmark	& \cmark$\star$	& Python	& scikit-optimize.space	\\
KTT~\cite{KTTSoftwareX}	& \cmark	& \cmark	& C++	& chain-of-trees	\\
ATF~\cite{ATF}	& \cmark	& \cmark	& C++	& chain-of-trees	\\
BaCO~\cite{BaCO2024}	& \cmark	& \xmark	& JSON	& chain-of-trees	\\
PyATF~\cite{pyATF}	& \cmark	& \cmark	& Python	& chain-of-trees	\\
Kernel Tuner	& \cmark	& \cmark	& Python	& CSP solver \\
\bottomrule
\end{tabular}
\end{table}

In the absence of constraints, the search space is defined as the Cartesian product of all possible combinations of all tunable parameter values. 
In brute-force search space construction, the approach is to simply iterate through all possible combinations and filter out combinations that violate the constraints.
This is reasonable for small search spaces, but becomes increasingly time-consuming as the search space size and number of constraints increase. 
For various auto-tuning applications, the number of valid configurations in the search space is orders of magnitude smaller than the Cartesian product size, causing the vast majority of generated and evaluated combinations to be discarded, wasting time and resources. 

\verb|ConfigSpace| and \verb|scikit-optimize.space| are both Python packages that implement functionality to represent multidimensional configuration spaces. 
Both approaches do not enumerate or store individual configurations, but instead provide an interface to generate random samples from the search space. 
As constraint resolution is not supported by \verb|scikit-optimize.space|, GPTune relies on an additional internal check on sampled points. %
While \verb|ConfigSpace| does allow users to specify constraints, called \textit{forbidden clauses} in \verb|ConfigSpace|, these constraints are again only checked after generating the sample point. 
The advantage of this approach is its simplicity and ability for uniform sampling over all points in the unconstrained Cartesian space.
However, the disadvantage of this approach is that constraints are not taken into account when generating samples, which has the same efficiency downsides as brute force search space construction.

Rasch et al.~\cite{ATF} introduced the chain-of-trees structure to represent constrained search spaces in constraint-based auto-tuning. 
Parameters are grouped based on interdependencies in the constraint functions, and each group is represented by a tree that encodes valid parameter combinations. 
These trees are then linked sequentially, forming a chain.
This structure can reduce redundancy by reusing shared subtrees, which may lead to lower memory usage compared to flat representations. 
Independent parameters are handled as single-parameter trees. %

In \cref{sec:searchspace_construction}, instead of adopting a customized solution for constraint-based auto-tuning such as chain-of-trees, we investigate the feasibility of using established methods with a robust mathematical foundation for efficient search space construction.

    \section{Application and Optimization of CSP Solvers}
\label{sec:searchspace_construction}

In this section, we discuss the design and implementation details of our novel method for efficiently constructing large search spaces for constraint-based auto-tuning.
\Cref{subsec:searchspace_construction_packages} examines various constraint-solving techniques to formalize the relation with auto-tuning and find a robust solver best suited to the problem context. 
\Cref{subsec:searchspace_construction_improvements_parser} describes the automatic optimization of user-defined constraints, followed by solver optimizations in \cref{subsec:searchspace_construction_improvements} to improve efficiency. 
Finally, \cref{subsec:searchspace_construction_searchspace_object} details how the resulting search space is represented and applied in auto-tuning frameworks.

\begin{lstfloat}[tb]
\begin{python}
p = Problem()
p.addVariables("block_size_x", [1,2,4,8,16]+[32*i for i in range(1,33)])
p.addVariables("block_size_y", [2**i for i in range(6)])
p.addConstraint(MinProd(32, ["block_size_x", "block_size_y"]))
p.addConstraint(MaxProd(1024, ["block_size_x", "block_size_y"]))
\end{python}
\vspace{-0.4cm}
\caption{Example of a \textit{python-constraint} problem definition.}
\label{code:python-constraint-example}
\end{lstfloat}

\subsection{Using Constraint Solvers in Auto-tuning} \label{subsec:searchspace_construction_packages}
The search space construction problem, where parameter values and constraints must be resolved to all valid combinations, can generally be encoded as a Boolean Satisfiability Problem (SAT)~\cite{biereHandbookSatisfiabilityVolume2009}, Satisfiability Modulo Theories (SMT)~\cite{barrett2008satisfiability}, or Constraint-Satisfaction Problem (CSP)~\cite{brailsfordConstraintSatisfactionProblems1999}. 
Among SAT, SMT, and CSP, the technique closest to auto-tuning search space construction is CSP. 
While SAT deals with Boolean variables and SMT extends SAT with theories like arithmetic or arrays, CSP solvers offer high-level abstractions suitable for modeling complex constraints that are otherwise difficult to efficiently express, making them better suited for modeling the complex relationships found in auto-tuning problems.

We can formalize the auto-tuning search space construction problem as a CSP defined by $\mathcal{P} = (X, D, C)$, where:
\begin{itemize}
  \item $X = \{x_1, x_2, \dots, x_n\}$ is a finite set of variables, each corresponding to a tuning parameter (e.g., block size, tile width).
  \item $D = \{D_1, D_2, \dots, D_n\}$ is a set of finite domains, where $D_i$ is the set of legal values for variable $x_i$.
  \item $C = \{c_1, c_2, \dots, c_m\}$ is a finite set of constraints, where each $c_j$ is a predicate over a subset of variables $\text{scope}(c_j) \subseteq X$. %
\end{itemize}

A solution to the auto-tuning search space is then a total assignment $ \mathcal{V}: X \rightarrow \bigcup D_i$ such that $\mathcal{V}(x_i) \in D_i$ for all $i$, and all constraints $c_j \in C$ are satisfied under $\mathcal{V}$.
\iflackofspace
The overall auto-tuning objective is to determine the optimal configuration as \\ $v^\star = \underset{v\in\mathcal{V}}{\text{arg max}} \, f_{H_j,I_k}(A_{i,v})$, where we have an application $A_i$ on a hardware platform $H_j$ for an input dataset $I_k$ to maximize performance $f_{H_j,I_k}(A_i)$ over the code variants in $\mathcal{V}$.
\fi

In addition, there are practical considerations when it comes to choosing a solver to build on. 
As we will implement our solver in the Python-based Kernel Tuner, the solver should be deployable in a Python environment. 
Notable options include \textit{OR-Tools}~\cite{llcOrtoolsGoogleORTools} and the Z3 solver in \textit{PySMT}~~\cite{teamPySMTSolveragnosticLibrary}, which provide highly expressive modeling capabilities and efficient solving algorithms. 
However, most solvers aim to find any solution, rather than all solutions, as required in the case of auto-tuning search space construction. 
To obtain all solutions, such solvers must iteratively find a solution, add this solution as an additional constraint, and look for the next solution until there are no solutions left~\cite{bjornerProgrammingZ32019}. 
If there are many solutions, as is commonly the case with auto-tuning problems, this can have a substantial negative impact on performance, as will be shown in~\cref{subsec:evaluation_synthetic}. 
A notable exception to this is \textit{python-constraint}~\cite{niemeyerPythonconstraintPythonconstraintModule}, as this is a CSP-based Python package capable of finding all solutions, which we will use as the basis of our implemented method. 

\subsection{Parsing Constraints} \label{subsec:searchspace_construction_improvements_parser}
Having formalized the auto-tuning search space construction to a CSP problem, we must now transform the Kernel Tuner constraints, such as in \cref{code:constraints}, to a format optimal for CSP solvers. 
To this end, we introduce a parser for constraints, which has three important benefits: 
to break down constraints into the smallest subsets of variables, 
to apply the more efficient specific constraints instead of generic functions where possible, 
and to provide optimal performance without requiring users of the auto-tuner to write constraints in a complex format that requires an understanding of how CSP solvers work. 

To address the latter benefit first, both CSP-solvers and auto-tuners have distinct interfaces consisting of specific function calls or a form of domain-specific language when defining the constraints. As seen before for auto-tuners in \cref{code:constraints}, an example of a \textit{python-constraint} problem definition is given in \cref{code:python-constraint-example}. 

However, as opposed to the users of CSP-solvers, auto-tuning users are generally not aware of the inner workings of the search space construction process and the specific constraints available that result in efficient resolution of the search space. 
Instead, we provide users the option to write their constraints as Python lambdas or the Python-evaluable string format, both seen in \cref{code:constraints}, which can then be optimized by our parser via Abstract Syntax Trees. 
This design has several benefits. It provides a familiar format for constraints, since Python is already used as the interface language. At the same time, our parser can rewrite these constraints, allowing the application of specific constraints instead of generic functions and the decomposition of constraints into subsets. 

\begin{figure}[bp]
    \centering
    \includegraphics[width=0.95\linewidth]{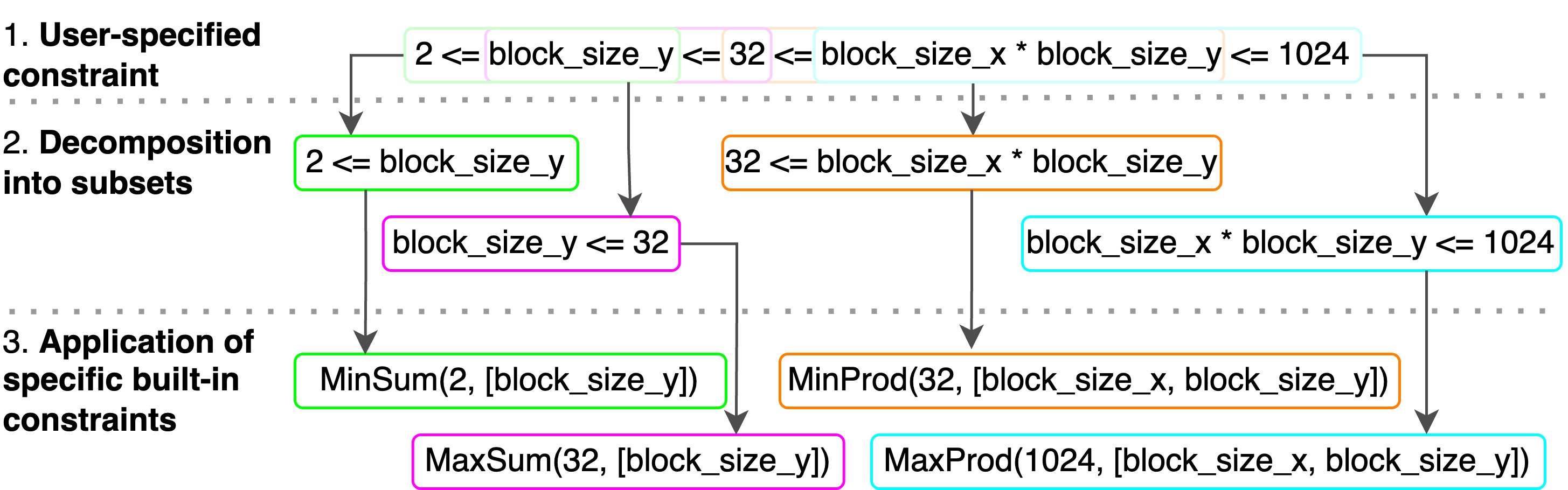}
    \vspace{-0.4cm}
    \caption{The optimization of a constraint via the parsing pipeline.}
    \Description[The constraints parsing pipeline given an example.]{The constraints parsing pipeline given an example.}
    \label{fig:constraints_parsing_pipeline_example}
\end{figure}

The automatic reduction of constraints is particularly important to obtain efficiency in practice, as users unfamiliar with the intrinsics of constraint solving, such as the users of auto-tuning frameworks, might write sub-optimal constraints. 
For example, consider the constraint \lstinline[language=Python]{2 <= block_size_y <= 32 <= block_size_x * block_size_y <= 1024}, 
slightly different from \cref{code:constraints,code:python-constraint-example}, where \lstinline[language=Python]{block_size_x}, \lstinline[language=Python]{block_size_y} are tunable parameters with numerical values. 
Constraints cannot be evaluated until values for the involved parameters are at least partially resolved, resulting in subpar performance in the case of compound statements like the given example, as it depends on the resolution of both parameters. 
This can be improved by automatically breaking down the composite constraint into multiple constraints with fewer variables where possible. 
This example can be decomposed as shown in Step 2 of \cref{fig:constraints_parsing_pipeline_example}, which allows partially resolved values for \lstinline[language=Python]{block_size_x} or \lstinline[language=Python]{block_size_y} to discard invalid configurations early. %

In addition, this automatic reduction enables the application of specific constraints, as is the case with the example, which can be represented with specific constraints as shown in Step 3 of \cref{fig:constraints_parsing_pipeline_example}. %
As will be further discussed in \cref{subsubsec:searchspace_construction_improvements_constraints}, the application of specific constraints can preemptively exclude values through preprocessing, resulting in an even more efficient construction. 

\subsection{Implementation of Optimizations} \label{subsec:searchspace_construction_improvements}
To obtain the level of performance required to construct auto-tuning search spaces efficiently, we implement several key improvements in various areas of the \textit{python-constraint} package: 
algorithmically (\cref{subsubsec:searchspace_construction_improvements_algorithm}), 
by extending constraints (\cref{subsubsec:searchspace_construction_improvements_constraints}), 
engineering (\cref{subsubsec:searchspace_construction_improvements_engineering}), 
and by tailoring output formats (\cref{subsubsec:searchspace_construction_improvements_output_formats}). 

\subsubsection{Algorithm} \label{subsubsec:searchspace_construction_improvements_algorithm}
We select and optimize a backtracking solver for finding all solutions rather than any solution. 
Shown simplified in \cref{alg:solver}, it maintains a dictionary of variable assignments and uses a stack-based approach for iterative backtracking, avoiding recursive function calls. 
For each selected variable, domain values are checked against the constraints. 
If a constraint is violated, the algorithm backtracks by restoring states until all possibilities are explored. %
We optimize this algorithm further by sorting the variables on the number of internal constraints, making it faster to find unassigned variables, and by reducing the number of sorts required.

\begin{algorithm}[tbp]
\scriptsize
\caption{Obtaining all valid configurations as $\mathcal{S}$ (simplified).} \label{alg:solver}
\KwIn{Variables $X=\{x_1,\dots,x_n\}$ with domains $D$, constraint set $C$}

\BlankLine
$\mathcal{A}\leftarrow\emptyset$; $\mathcal{S}\leftarrow\emptyset$ \Comment*{current (partial) assignment; solution set}
$\pi\leftarrow\mathrm{SortVariables}(X)$ \Comment*{sorted on number of constraints involved}
$B\leftarrow\langle(\pi,\mathcal{A})\rangle$ \Comment*{backtrack stack of states}

\While(\tcp*[f]{main search loop}){$B\neq\emptyset$}{
    $(\pi,\mathcal{A})\leftarrow B.\mathrm{pop}()$\;
    
    \uIf(\tcp*[f]{add to solutions if all variables assigned}){$\forall x\in \pi: x\in\mathcal{A}$}{
        $\mathcal{S}\leftarrow \mathcal{S}\cup\{\mathcal{A}\}$; \textbf{continue}\;
    }
    
    $x\leftarrow \mathrm{NextUnassigned}(\pi,\mathcal{A})$\;
    
    \ForEach(\tcp*[f]{try all values for $x$}){$v\in D(x)$}{
        $\mathcal{A}[x]\leftarrow v$\;
        
        \lIf{$\mathrm{CheckConstraints}(\mathcal{A},C)$}{
            $B.\mathrm{push}\bigl(\pi,\mathcal{A}\bigr)$
        }
        
        $\mathcal{A}.erase(x)$ \Comment*{undo and try next value}
    }
}
\Return{$\mathcal{S}$}\;
\end{algorithm}

\subsubsection{Constraints} \label{subsubsec:searchspace_construction_improvements_constraints}
We expand and improve built-in specific constraints to optimize constraint operators commonly used in auto-tuning.  
By applying knowledge of the operation, their efficiency can be improved over generic functions. 
For example, given a constraint where $p \cdot q > 0$, we know to ignore all cases where $(p \leq 0) \xor (q \leq 0)$. 
We add \textit{MaxProduct} and \textit{MinProduct} constraints as they are commonly used in auto-tuning constraints (e.g. a maximum product of block sizes). %
We also improve and add preprocessing steps to the various existing constraints, such as \textit{MaxSum} and \textit{MinSum}. 
All specific constraints are precompiled for further efficiency gains. 

However, not all constraints can be expressed as built-in constraints, for example when using an operation that is not as common. 
Such cases are parsed to \textit{Function} constraints, which we have optimized by employing function rewriting and dynamic runtime compilation, as the one-off expense of compilation to bytecode is offset by the many times a \textit{Function} constraint is usually executed.

\subsubsection{Employing C-extensions} \label{subsubsec:searchspace_construction_improvements_engineering}
In general, C and similar languages outperform Python in terms of execution speed~\cite{pythonVSC,comparingSixLanguages}. To attain this level of performance without losing the flexibility and user-friendliness of Python~\cite{pythonVSC++usability}, we employ C-extensions. 
We transpile the codebase from Python to C-code using Cython~\cite{behnelCythonBestBoth2011}, which is then compiled into Python-importable C-extensions. 
We added type hints where possible to aid in compilation. 

\subsubsection{Output Formats} \label{subsubsec:searchspace_construction_improvements_output_formats}
We implement various output formats to avoid expensive rearrangements to different formats. 
Expensive rearrangement of the structure in which solutions are output by the solver is mitigated by providing output formats that are close to the internal representation, further described in \cref{subsec:searchspace_construction_searchspace_object}.

\subsection{Search Space Representation} \label{subsec:searchspace_construction_searchspace_object}
With the efficient construction of search spaces implemented in \textit{python-constraint}, we consider how this is represented and applied in auto-tuning frameworks for a comprehensive approach. 

After the search space construction, optimization algorithms use the information obtained in the construction step to select configurations.
Instances of this are obtaining the true bounds of the search space to use balanced initial sampling methods or the selection of valid neighbors that have not been evaluated yet. 

This highlights a key advantage of our method over the dynamic approaches discussed in \cref{sec:related_work}. 
Important search space characteristics, such as the true parameter bounds, can guide optimization algorithms more effectively and facilitate the use of stratified sampling techniques, such as Latin Hypercube Sampling~\cite{willemsenBayesianOptimizationAutotuning2021}.
However, these characteristics can not be reliably used in dynamic approaches, as a resolved search space is required. 
Furthermore, randomized sampling is inherently biased to the sparser parts of the chain-of-trees, although this has been addressed by BaCO~\cite{BaCO2024}. 
Moreover, selecting valid neighbors of configurations as extensively used by various optimization algorithms is potentially expensive. 

Instead, we fully resolve the search space before starting the tuning process, with a minimal impact on the total execution time, to incorporate the full information of the search space in the initial sampling and optimization algorithms. 
As operations such as sampling and finding valid neighbors are commonly used in auto-tuning, it can be useful to provide an abstract representation of the search space that implements these operations, providing various views and mappings on the configurations in the search space. 

We have implemented this in Kernel Tuner as the \textit{SearchSpace} class, which takes the tunable parameters and constraints based on the user specification, constructs the search space using our implementation, and provides various representations and operations on the resulting search space.  
The \textit{SearchSpace} class has multiple internal representations for varying purposes, such as hash- and index-based for efficient lookups. 
Externally, it provides a single interface for all search space-related operations, promoting reuse. 
For example, the mutation step in the \textit{genetic algorithms} optimization algorithm requires selecting only valid neighbors within a certain Hamming distance. This, along with other neighbor selection algorithms, is implemented in the \textit{SearchSpace} class and can be indexed before running the algorithm, improving overall performance.

    \section{Evaluation}\label{sec:evaluation}

In this section, we evaluate the advancements presented in \cref{sec:searchspace_construction} to determine their scalability and performance impact.
First, we discuss how we compare against the current state-of-the-art solvers in \cref{subsec:evaluation_compared_frameworks}.
We then evaluate the solvers on a large collection of synthetically generated search spaces with varying characteristics to assess scalability differences in \cref{subsec:evaluation_synthetic}. 
Following this, we evaluate the solvers on a variety of real-world applications to assess the performance improvement in \cref{subsec:evaluation_real-world}. 
Finally, we validate the practical impact of our method on the entire auto-tuning pipeline in \cref{subsec:evaluation_tuning_test}.

The evaluations in this work are performed on the sixth generation DAS \href{https://www.cs.vu.nl/das/clusters.shtml}{VU-cluster}~\cite{DASMediumScaleDistributedSystem} using an NVIDIA A100 GPU node. 
The GPU is paired with a 24-core AMD EPYC-2 7402P CPU, 128 GB of memory, and running Rocky Linux 4.18. 
For all tests, the results of each solver were validated against a brute-force solution of the search space. 
\ifdoubleblind
Our evaluation implementation is publicly available. 
\else
Our evaluation implementation is \href{https://github.com/fjwillemsen/kernel_tuner_paper}{publicly available}.
\fi

\subsection{Comparison against state-of-the-art} \label{subsec:evaluation_compared_frameworks}
To provide additional reference on the performance in this evaluation, we compare the results to the state-of-the-art in auto-tuning search space construction, the chain-of-trees of Auto-Tuning Framework (ATF). 
ATF has two independent implementations, in C++~\cite{ATF} and Python~\cite{pyATF}, both of which we use in this evaluation to compare our method. 
The C++ version available as of August 2024 with Python bindings is used and denoted as \textit{ATF} in the results. 
The Python version called \textit{pyATF} is used at version 0.0.9, the latest version at the time of writing.

Due to the large number of search spaces used in this evaluation, it is not feasible to write each of these search space definition files by hand for both ATF implementations, and we have instead written parsers that define the ATF search space files from an abstract definition of the search spaces. 
Both implementations of ATF have a notation that combines the definition of tunable parameters, values, and constraints into one statement, as seen in \cref{code:constraints}. 
Hence, ATF constraints can only reference tunable parameters that have been previously defined. 
To provide search space definitions that are compliant with this ATF format, the parsers account for the parameter-constraint order relation and convert to built-in ATF types, such as intervals, where applicable. 
To reflect the user experience as accurately as possible, the search space file compilation time is included in the total construction time. 
The C++ version of ATF and search space files is compiled with GCC 12.4.0 using the optimization commands recommended by the ATF documentation. 

In addition, we compare with PySMT version 0.9.6, using the Z3 solver developed by Microsoft for software verification and analysis \cite{Z3solver}.
This allows for evaluation of scalability differences for solvers that do not support enumerating all solutions, as discussed in \cref{subsec:searchspace_construction_packages}.
As with ATF, we developed a custom parser that employs PySMT-specific operations where applicable.

\subsection{Synthetic Tests} \label{subsec:evaluation_synthetic}
To understand how search space characteristics influence construction time and scaling of solvers, we evaluate on synthetic tests. 

\subsubsection{Experimental setup} \label{subsubsec:evaluation_synthetic_setup}
We generate a set of search spaces with a varying number of dimensions (between 2 and 5), target Cartesian sizes (with $\{1\times 10^{4}, 2\times 10^{4}, 5\times 10^{4}, 1\times 10^{5}, 2\times 10^{5}, 5\times 10^{5}, 1\times 10^{6}\}$), and number of constraints (between 1 and 6). 
While these arbitrary parameters result in search spaces that are not as large and do not have as many tunable parameters as the real-world search spaces evaluated on in \cref{subsec:evaluation_real-world}, the goal of these in total 78 synthetic search spaces is to gain insight into which solver provides good scalability across the variations in these factors.

\begin{figure}[tbp]
    \centering
    \includegraphics[width=0.975\linewidth]{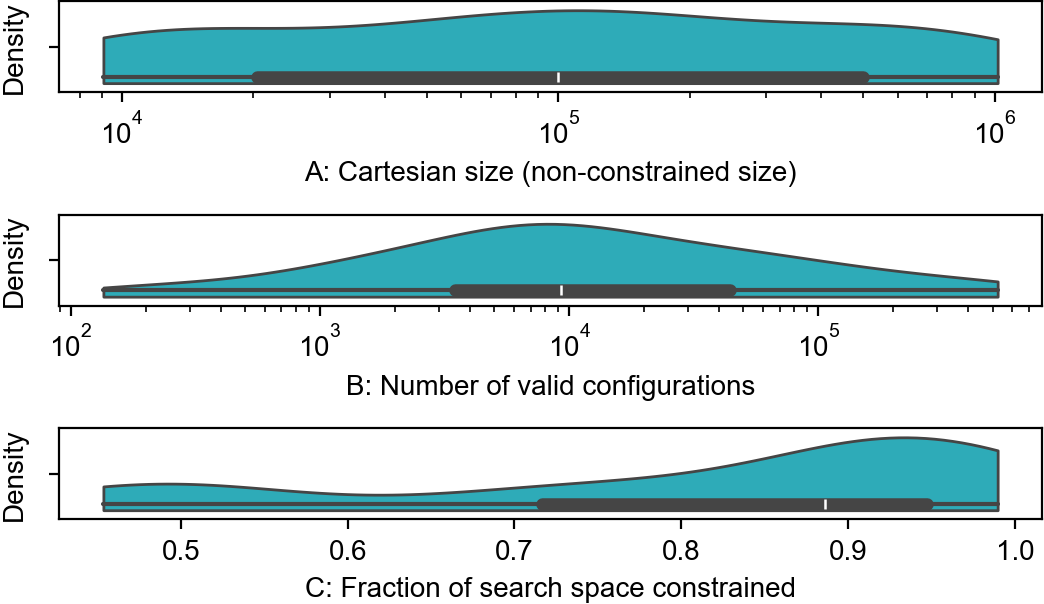}
    \vspace{-0.4cm}
    \caption{Density of three characteristics of the 78 synthetic search spaces. Black bottom bar marks the interquartile range and the white line the median.}
    \Description[Characteristics of the 78 synthetic search spaces.]{Characteristics of the 78 synthetic search spaces.}
    \label{fig:searchspaces_synthetic_characteristics}
\end{figure}

\begin{figure*}[tbp]
    \centering
    \includegraphics[width=1.0\textwidth]{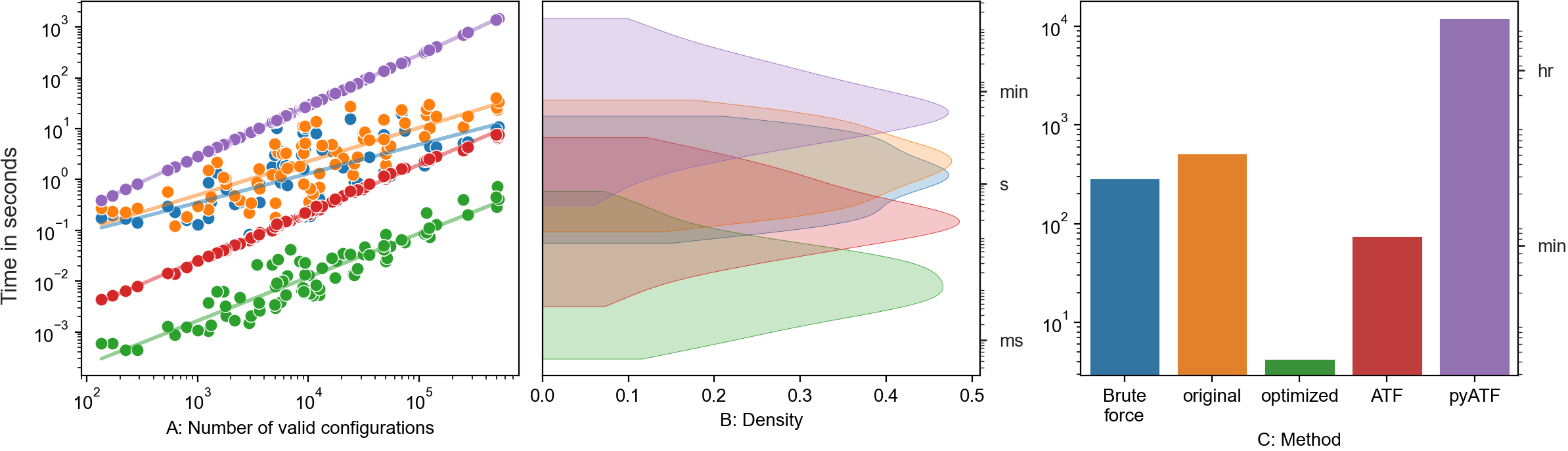}
    \vspace{-0.7cm}
    \caption{Search space construction performance on synthetic tests. Lower times are better. Colors correspond to \Cref{fig:results_synthetic}C barplot methods. Each plot provides a different view of the same data, with A and B showing the performance on individual search spaces, and C showing the sum of all search spaces.}
    \Description[search space construction performance]{search space construction performance on synthetic tests. Lower times are better.}
    \label{fig:results_synthetic}
\end{figure*}

Given a Cartesian size, a number of dimensions, and a number of constraints, we want to generate a synthetic search space. 
To prevent an unfair advantage to solvers optimized for a limited number of dominant dimensions, the number of values per dimension $v$ is kept approximately uniform. 
This is done by first determining the number of values per dimension as $v = s^{\frac{1}{d}}$, where $s$ is the desired Cartesian size and $d$ is the desired number of dimensions. 
For each of the dimensions, a linear space with $v$ number of elements is instantiated. 
Given a non-integer value of $v$, this is rounded to an integer for all but the last dimension, where $v$ is rounded contradictory (e.g. $5.8 \rightarrow 5$, $5.2 \rightarrow 6$) to be closer to the desired Cartesian size. 
A list of constraints involving a variety of operations is generated for each combination of dimensions, which are randomly chosen up to the desired number of constraints. 

\Cref{fig:searchspaces_synthetic_characteristics} shows the distribution of the resulting 78 search spaces for three characteristics.
\Cref{fig:searchspaces_synthetic_characteristics}A shows the actual Cartesian size, representing the total number of possible configurations before constraints are applied, which corresponds to the set of target values. 
\Cref{fig:searchspaces_synthetic_characteristics}B depicts the number of valid configurations remaining after constraints are enforced, resulting in an approximately bell-shaped curve. The number of valid configurations is on average one order of magnitude below the Cartesian size. 
Finally, \Cref{fig:searchspaces_synthetic_characteristics}C displays the fraction of sparsity of the search space, i.e., the fraction of non-valid configurations relative to the Cartesian size. 
Though the fraction of constrained configurations is skewed toward higher values, indicating a propensity towards sparsity, a wide range of variations in sparsity is present. 

\subsubsection{Results} \label{subsubsec:evaluation_synthetic_results}
The performance of the evaluated methods on these synthetic search spaces is displayed in various plots in \Cref{fig:results_synthetic}, where the colors used correspond to the colors of the methods in the \cref{fig:results_synthetic}C barplot. 
To determine the impact of the optimizations described in \cref{subsec:searchspace_construction_improvements}, the \textit{original} method denotes the use of vanilla \textit{python-constraint} before the optimizations, whereas our \textit{optimized} method includes the optimizations of \cref{subsec:searchspace_construction_improvements}. 

\Cref{fig:results_synthetic}A shows a clear positive correlation between the number of valid configurations and the execution time across all methods, with a roughly linear trend on the log-log scale, suggesting a power-law relationship. 
We overlay a log-log linear regression to further investigate the scaling of each method, where a lower slope indicates better scaling towards larger search spaces in the number of valid configurations, and a slope of 1 indicates linear scaling. All methods have a highly significant linear fit with a p-value $\ll 0.05$. 

For the methods \textit{ATF} and \textit{pyATF}, we observe approximately linear scaling, with slopes of 0.938 and 0.999, respectively. 
The \textit{original} unoptimized python-constraint and \textit{brute force} methods appear to perform similarly to each other and exhibit good scaling with slopes of 0.663 and 0.571, respectively. 
While they are outperformed by \textit{ATF} on these search spaces, the difference in scaling means both methods appear to soon outperform \textit{ATF} on larger search spaces, which we extrapolate to be at $\sim 1.193 \cdot 10^6$ and $\sim 4.493 \cdot 10^7$ valid configurations respectively. 
It is important to note that this difference in scaling is expected, given that brute-forcing will perform relatively better the denser a search space is (i.e. many valid configurations relative to the Cartesian size of the search space) as it will check the constraints on all configurations, where the chain-of-trees is optimized towards sparse search spaces. 
Our \textit{optimized} method is consistently the fastest, always constructing the search space within less than a second, and exhibits adequate scaling with a slope of 0.860. As based on this data our \textit{optimized} method would not be overtaken by the \textit{brute force} and \textit{original} methods until $\sim 1.120 \cdot 10^{11}$ and $\sim 3.892 \cdot 10^{15}$ valid configurations respectively, well beyond the size of these synthetic search spaces, we expect our method to perform best overall.

\Cref{fig:results_synthetic}B presents the performance as a continuous probability density curve using a kernel density estimate (KDE). 
As expected due to their practically linear scaling, ATF and pyATF demonstrate wide variance in performance. 
Our \textit{optimized} solver consistently achieves the lowest execution times, with several orders of magnitude better performance compared to the other solvers.

\Cref{fig:results_synthetic}C summarizes the performance of each solver over all search spaces. %
It is remarkable that \textit{pyATF} takes considerably longer than the \textit{brute-force} method on these search spaces, which might be due to how optimized the chain-of-trees approach is to highly sparse search spaces.
It is also noteworthy that our \textit{optimized} method outperforms the \textit{original} unoptimized implementation of \textit{python-constraint} by several orders of magnitude, demonstrating the advantage of our optimizations. 
Our \textit{optimized} method achieves a 96x speedup over the \textit{brute-force} method (4.75 seconds versus 455.3 seconds), a 16x speedup over \textit{ATF}, and a 2547x speedup over \textit{pyATF}.

\begin{figure}[tbp]
    \centering
    \vspace{-0.3cm}
    \includegraphics[width=1.0\linewidth]{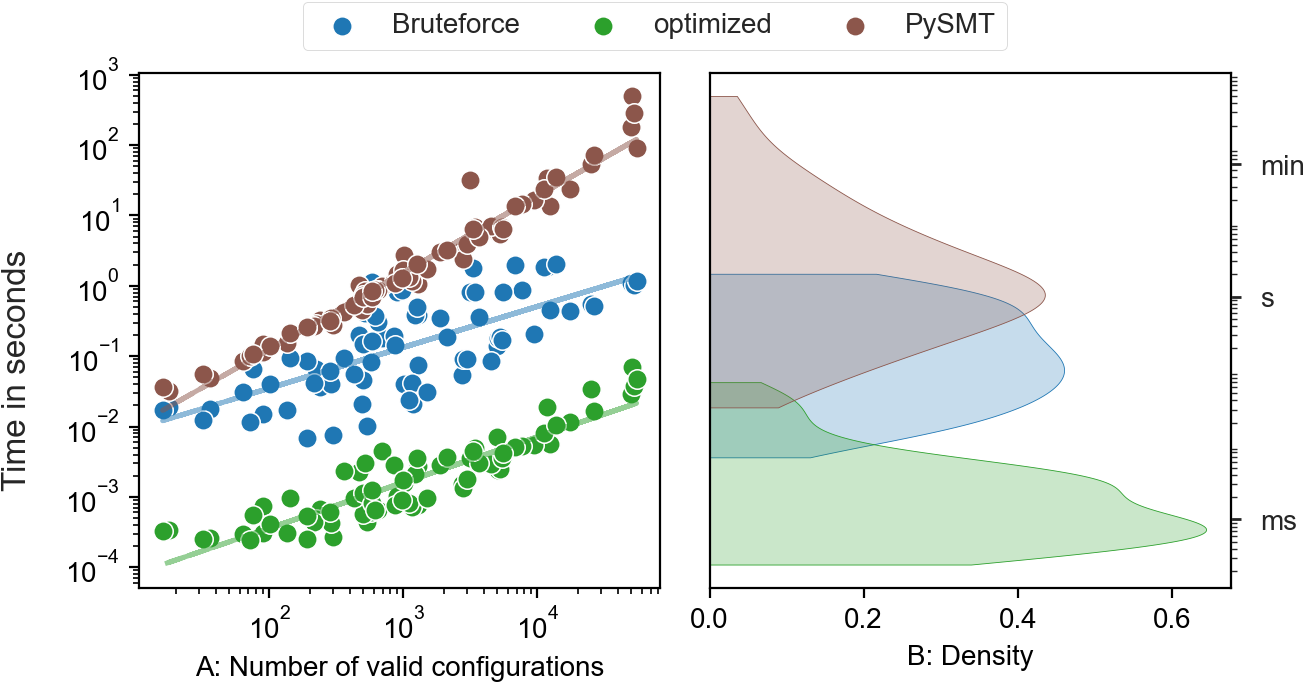}
    \vspace{-0.8cm}
    \caption{Search space construction performance of PySMT on synthetic tests.} %
    \Description[search space construction performance synthetic with PySMT]{search space construction performance on synthetic tests, including PySMT.}
    \label{fig:results_synthetic_pysmt}
\end{figure}

As described in \cref{subsec:searchspace_construction_packages}, a traditional solver without support for finding all solutions requires adding the previous solution as a constraint and iterating over the solutions until all solutions have been found. To demonstrate the lack of scalability of such a solver, \cref{fig:results_synthetic_pysmt} compares \textit{PySMT} using the Microsoft Z3 solver to the \textit{brute-force} method and our \textit{optimized} method. 
To make executing this experiment feasible, we reduce the size of the generated synthetic search spaces by one order of magnitude in this experiment. 
As seen in \cref{fig:results_synthetic_pysmt}, PySMT performs poorly relative to both brute force and our optimized method. As expected, this difference increases as the number of valid configurations increases, demonstrating the infeasibility of this approach when many valid configurations are present. 
Despite the reduced search space sizes, PySMT with the Z3 solver still takes nearly a thousand seconds on the largest search spaces, whereas the brute-force solver takes about ten seconds. 
Our optimized solver vastly outperforms PySMT, taking about as long to solve the largest search spaces as PySMT with the Z3 solver takes to solve the smallest search spaces. 
In fact, the PySMT solver exhibits superlinear scaling with a slope of 1.090, as opposed to the slope of 0.649 of our optimized method.
As it is infeasible to evaluate the large search spaces of the selected real-world applications, PySMT with the Z3 solver will not be included in the remainder of the evaluation.

\subsection{Real-world Applications} \label{subsec:evaluation_real-world}

\begin{table*}[tbp]
    \centering
    \scriptsize
    \caption{Overview of the basic characteristics of the real-world search spaces and the mean values for each of the columns.}
    \label{tab:searchspaces_real_world_overview}
    \vspace{-0.3cm}
    \begin{tabularx}{\linewidth}{|l|X|X|X|X|X|X|X|X|}
        \hline
        
        \textbf{Name} & \textbf{Cartesian size} & \textbf{Constraint size} & \textbf{Number of parameters (dimensions)} & \textbf{Number of constraints} & \textbf{Avg. unique parameters per constraints} & \textbf{Range of number of values per parameter} & \textbf{\% of configurations in Cartesian size} & \textbf{Avg. number of constraint evaluations required}  \\
        \hline
        Dedispersion & 22272 & 11130    & 8 & 3     & 2 & 1 - 29 & 49.973     & 33414 \\\hline
        ExpDist & 9732096 & 294000      & 10 & 4    & 2 & 1 - 11 & 3.021      & 23889240 \\\hline
        Hotspot & 22200000 & 349853     & 11 & 5    & 3.8 & 1 - 37 & 1.576    & 65900294 \\\hline
        GEMM & 663552 & 116928          & 17 & 8    & 3.25 & 1 - 4 & 17.622   & 2576736 \\\hline
        MicroHH & 1166400 & 138600      & 13 & 8    & 2.375 & 1 - 10 & 11.883 & 4763700 \\\hline
        ATF PRL 2x2 & 36864 & 1200      & 20 & 14   & 2.429 & 1 - 3 & 3.255   & 268680 \\\hline
        ATF PRL 4x4 & 9437184 & 10800   & 20 & 14   & 2.429 & 1 - 4 & 0.114   & 70708680 \\\hline
        ATF PRL 8x8 & 2415919104 & 48720 & 20 & 14  & 2.429 & 1 - 8 & 0.002   & 18119076600 \\\hline
        \hline
        \textit{Mean} & \textit{307322534} & \textit{121403} & \textit{14.875} & \textit{8.75}  & \textit{2.589} & \textit{1} - \textit{13.25} & \textit{10.93} & \textit{2285902168} \\\hline
    \end{tabularx}
\end{table*}

To evaluate solver performance on the search spaces of real-world applications, we select the three largest search spaces in the Benchmark suite for Auto-Tuners (BAT)~\cite{BenchmarkingSuiteKerneltuners}. 
These are \textit{Dedispersion}, \textit{Hotspot}, and \textit{ExpDist}. 
In addition, we use the relatively large search spaces of the \textit{MicroHH} computational fluid dynamics kernel~\cite{MicroHH2017}, as well as the commonly used General Matrix Multiplication kernel \textit{(GEMM)}~\cite{CLBlast2018}. 
To provide reference points for a fair comparison to ATF, the Probabilistic Record Linkage (\textit{PRL}) kernel used in the chain-of-trees evaluation~\cite{searchspaceATF} is used as well, resulting in three additional search spaces for a total of eight real-world search spaces. 

The characteristics of the real-world search spaces are displayed in \cref{tab:searchspaces_real_world_overview}, where the rightmost column denotes the average number of constraint evaluations that are required to brute-force solve a search space. For each combination in the Cartesian product, all constraints need to be evaluated until the combination is rejected or all constraints have been evaluated. Hence, assuming uniform probability among the constraints, the average number of constraint evaluations can be calculated by taking the average of the best case (the first constraint rejects the combination) and worst case (the last constraint rejects the combination), and adding all valid combinations that are never rejected. Given a search space $S$, let $S_i$ be the set of non-valid combinations, $S_v$ the set of valid combinations, and $S_c$ the set of constraints, the average number of constraint evaluations can be calculated as $\frac{|S_i|+|S_i| \cdot |S_c|}{2}+|S_v|$. 
Descriptions of each of the kernels and their search spaces are given in \cref{subsubsec:evaluation_setup_kernel_dedispersion,subsubsec:evaluation_setup_kernel_expdist,subsubsec:evaluation_setup_kernel_hotspot,subsubsec:evaluation_setup_kernel_gemm,subsubsec:evaluation_setup_kernel_microhh,subsubsec:evaluation_setup_kernel_atf_prl}, before the results are discussed in \cref{subsubsec:evaluation_real-world_results}.

\subsubsection{Dedispersion} \label{subsubsec:evaluation_setup_kernel_dedispersion}
\iflackofspace
The Dedispersion kernel introduced in~\cite{scloccoAutoTuningDedispersionManyCore2014,sclocco2020amber} and used in~\cite{BenchmarkingSuiteKerneltuners} is designed to compensate for the time delay experienced by radio waves as they propagate through space. This delay occurs due to the frequency-dependent dispersion of the signal. By applying a specific dispersion measure (DM) and reversing the dispersion effect, the kernel reconstructs the original signal.
During the iteration over frequency channels, threads process multiple time samples and dispersion measures in parallel.
\fi
Comparing the Dedispersion search space to the other evaluated search spaces of \cref{tab:searchspaces_real_world_overview}, the resulting search space is the smallest in Cartesian size, but as it has the highest percentage of valid configurations at nearly 50~\%, it is not the smallest in number of valid configurations.

\subsubsection{ExpDist} \label{subsubsec:evaluation_setup_kernel_expdist}
\iflackofspace
The ExpDist kernel described in~\cite{BenchmarkingSuiteKerneltuners} is utilized in a localization microscopy application that performs template-free particle fusion by integrating multiple observations into a single super-resolution reconstruction~\cite{heydarian2018template}. During the registration process, the ExpDist kernel is repeatedly invoked to evaluate the alignment of two particles.
The algorithm exhibits quadratic complexity with respect to the number of localizations per particle, making it highly computationally intensive. 
\fi
The resulting search space is the second-most sparse of the real-world search spaces in \cref{tab:searchspaces_real_world_overview}.

\subsubsection{Hotspot} \label{subsubsec:evaluation_setup_kernel_hotspot}
Previously discussed in \cref{sec:background}, the Hotspot kernel of~\cite{BenchmarkingSuiteKerneltuners} is part of a thermal simulation application to estimate the temperature of a processor by considering its architecture and simulating power currents. Through an iterative process, the kernel solves a set of differential equations. The inputs to the kernel consist of power and initial temperature values, while the output is a grid displaying average temperature values across the chip. %
It is interesting to note that the Hotspot search space is the largest in number of valid configurations, second-largest in Cartesian size, and has the highest number of values for a single parameter.

\subsubsection{MicroHH} \label{subsubsec:evaluation_setup_kernel_microhh}
The computational fluid dynamics kernel of~\cite{MicroHH2017} is used for weather and climate modeling, specifically for the simulation of turbulent flows in the atmospheric boundary layer. 
In this case, we use the search space resulting from the auto-tunable GPU implementation of the \verb|advec_u| kernel with extended parameter values as specified in the source of~\cite{heldensKernelLauncherLibrary2023}. %
Looking at \cref{tab:searchspaces_real_world_overview}, it is notable that the MicroHH search space is the closest to the mean values of all search spaces in the number of parameters, number of constraints, and percentage of configurations. It is also second-closest in constraint size and number of values per parameter, making it perhaps the most average search space in our set of tests.

\subsubsection{GEMM} \label{subsubsec:evaluation_setup_kernel_gemm}
Generalized dense matrix-matrix multiplication is a fundamental operation in the BLAS linear algebra library and widely used across various application domains. 
Known for its high performance on GPU hardware, GEMM frequently serves as a benchmark in studies of GPU code optimization~\cite{CLTune,li2009note,pruning}. 
In this evaluation, we use the GEMM kernel of CLBlast~\cite{CLBlast2018}, a tunable OpenCL BLAS library. 
\iflackofspace
GEMM is implemented as the multiplication of two matrices ($A$ and $B$); $C = \alpha A \cdot B + \beta C$, where $\alpha$ and $\beta$ are constants and $C$ is the output matrix. 
\fi
The dimensions of all three matrices are set to $4096 \times 4096$, resulting in a dense search space. %

\subsubsection{ATF PRL} \label{subsubsec:evaluation_setup_kernel_atf_prl}
The Probabilistic Record Linkage (PRL) kernel used in \cite{searchspaceATF} is a parallelized implementation of an algorithm that is commonly used in data mining to identify data records referring to the same real-world entity. 
In this kernel, the input sizes determine the size of the search space. As shown in \cref{tab:searchspaces_real_world_overview}, the brute-force resolution of this search space with input sizes $8x8$ requires $1.8119 \times 10^{10}$ constraint evaluations on average, which took $\sim$27 hours to execute. %
As an input size of $16x16$ would require $4.639 \times 10^{12}$ constraint evaluations on average, it is not feasible to brute force beyond the $8x8$ input size. %
Because the brute-forced solution is used for validation and serves as a reference point in the performance comparisons, we use the search spaces resulting from the ATF PRL kernel with input sizes $2x2$, $4x4$, and $8x8$. It is notable that while the $8x8$ search space results in the largest Cartesian size of the set, the ATF PRL search spaces are very sparse.

\begin{figure*}[tbp]
    \centering
    \includegraphics[width=1.0\textwidth]{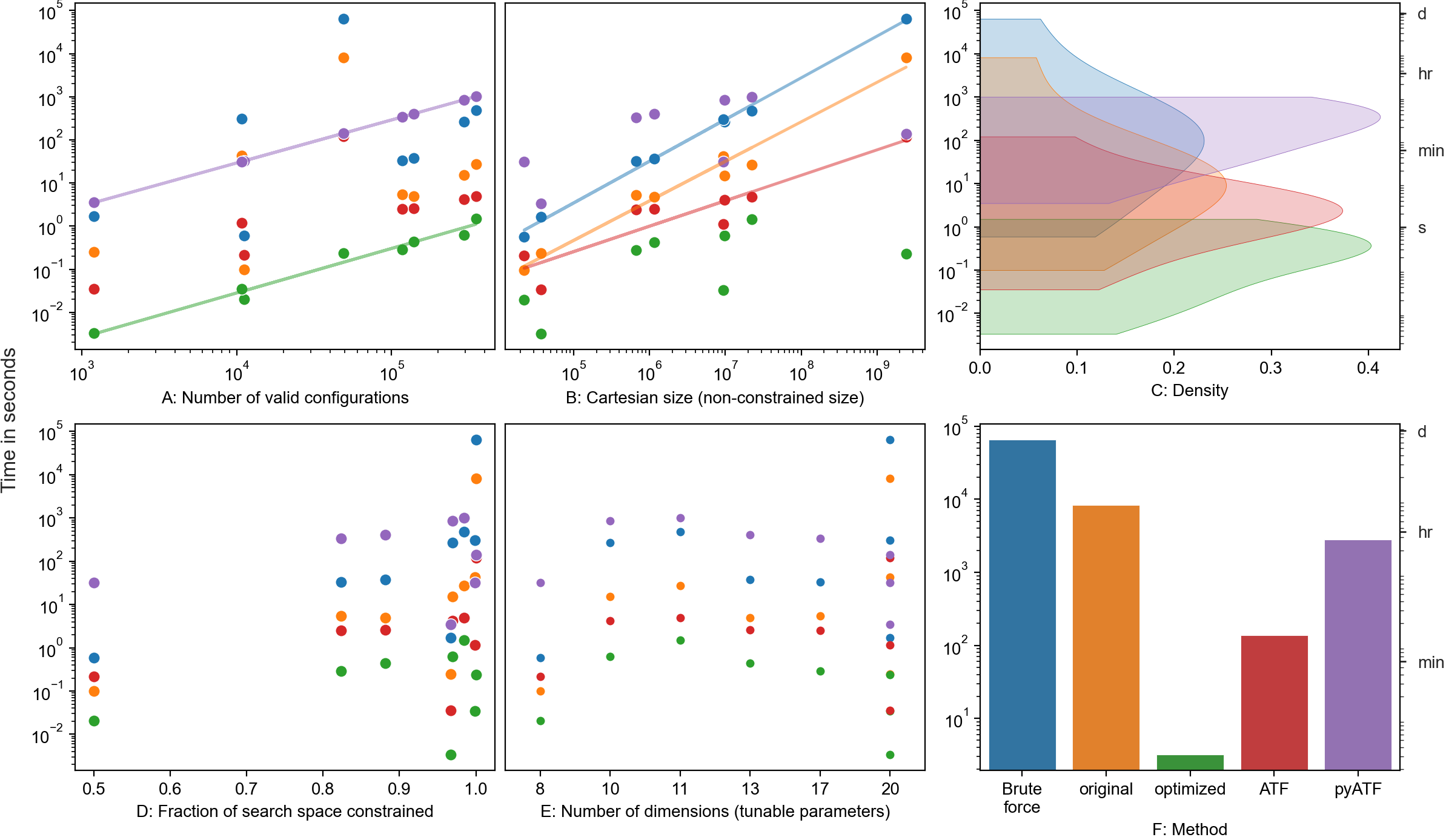}
    \vspace{-0.7cm}
    \caption{Search space construction performance on real-world tests. Lower times are better. Colors correspond to \Cref{fig:results_realworld}F barplot methods. Each plot provides a different view of the same data; plots A-E show individual performance relative to a search space characteristic, and plot F shows the sum of all search spaces.}
    \Description[Real-world search space construction performance]{Search space construction performance on real-world tests. Lower times are better.}
    \label{fig:results_realworld}
\end{figure*}

\subsubsection{Results} \label{subsubsec:evaluation_real-world_results}
\Cref{fig:results_realworld} presents the search space construction performance across the eight real-world benchmarks for the five different constraint solver methods, as before in \cref{subsec:evaluation_synthetic}. %

\Cref{fig:results_realworld}A and \cref{fig:results_realworld}B illustrate the relationship between search space size and solver performance, with a log-log linear regression overlayed where significant (p-value $\leq 0.05$), as in \cref{subsubsec:evaluation_synthetic_results}. 
In general, larger constrained search spaces (A) and Cartesian sizes (B) result in increased search times, as previously observed in \cref{fig:results_synthetic}. %
For the \textit{ATF}, \textit{original}, and \textit{brute-force} methods, the significant scaling trend is along the Cartesian size of \cref{fig:results_realworld}B, whereas for \textit{pyATF} and our \textit{optimized} method, this is on the number of valid configurations in \cref{fig:results_realworld}A. 
Our \textit{optimized} method achieves the lowest execution times across all search space sizes, demonstrating its efficiency, and is the only solver that consistently outperforms the other methods. 

\Cref{fig:results_realworld}C visualizes the distribution of execution times, providing an indication of the average performance and variability. 
It is interesting to observe that while the \textit{original} python-constraint method is one order of magnitude faster than the \textit{brute-force} method, both methods have very similar distributions, as seen before in \cref{fig:results_synthetic}B. 
A clear trend emerges from this plot, where our \textit{optimized} method has the least variability and is the only solver constructing the search spaces in the sub-second domain. 

In \cref{fig:results_realworld}D, the relation between how constrained a search space is and solver performance is displayed. 
\textit{ATF} and \textit{pyATF} performance appears to be strongly influenced by the sparsity, as for fraction $> 0.9$ \textit{ATF} performance is substantially better than the \textit{original} solver, in contrast to $\leq 0.9$, where at fraction $\simeq 0.5$ even the unoptimized \textit{original} python-constraint outperforms \textit{ATF}. 

The number of tunable parameters displayed in \cref{fig:results_realworld}E does not appear to have as much of an impact on performance as the other plots discussed. 
Nevertheless, \cref{fig:results_realworld}E is useful to discern the individual search spaces based on the number of parameters described in \cref{tab:searchspaces_real_world_overview}. 
For instance, it can be noted that the performance difference between our optimized method and all other methods appears to be relatively stable, even for the ATF PRL search spaces, as can be discerned by the number of tunable parameters, where the three ATF search spaces have 20 tunable parameters.

Finally, \cref{fig:results_realworld}F summarizes the total time taken by each solver. 
The brute-force approach is the least performant, taking nearly a full day to resolve the eight search spaces. 
Although the \textit{original} python-constraint solver is faster than brute force, our \textit{optimized} solver achieves a $\sim$2643x speedup over it, demonstrating the efficiency of our optimizations.
While ATF and pyATF achieve intermediate performance levels, it must be noted that pyATF only outperforms \textit{brute force} and \textit{original} because it does so on the two largest PRL search spaces, which have a disproportionate effect on the summed time - on all other search spaces, pyATF is outperformed by both methods, and ATF is not consistently better than \textit{original} either. 
The optimized solver consistently and considerably outperforms all others: our \textit{optimized} method achieves a $\sim$20643x overall speedup over the brute-force method (3.16 seconds versus 65230.47 seconds), $\sim$44x over ATF, and $\sim$891x over pyATF. 

\subsection{Overall impact in practice} \label{subsec:evaluation_tuning_test}
To conclude this evaluation, we evaluate the impact of the search space construction method on the overall auto-tuning process. %

\begin{figure}[tbp]
    \centering
    \vspace{-0.4cm}
    \includegraphics[width=1.0\linewidth]{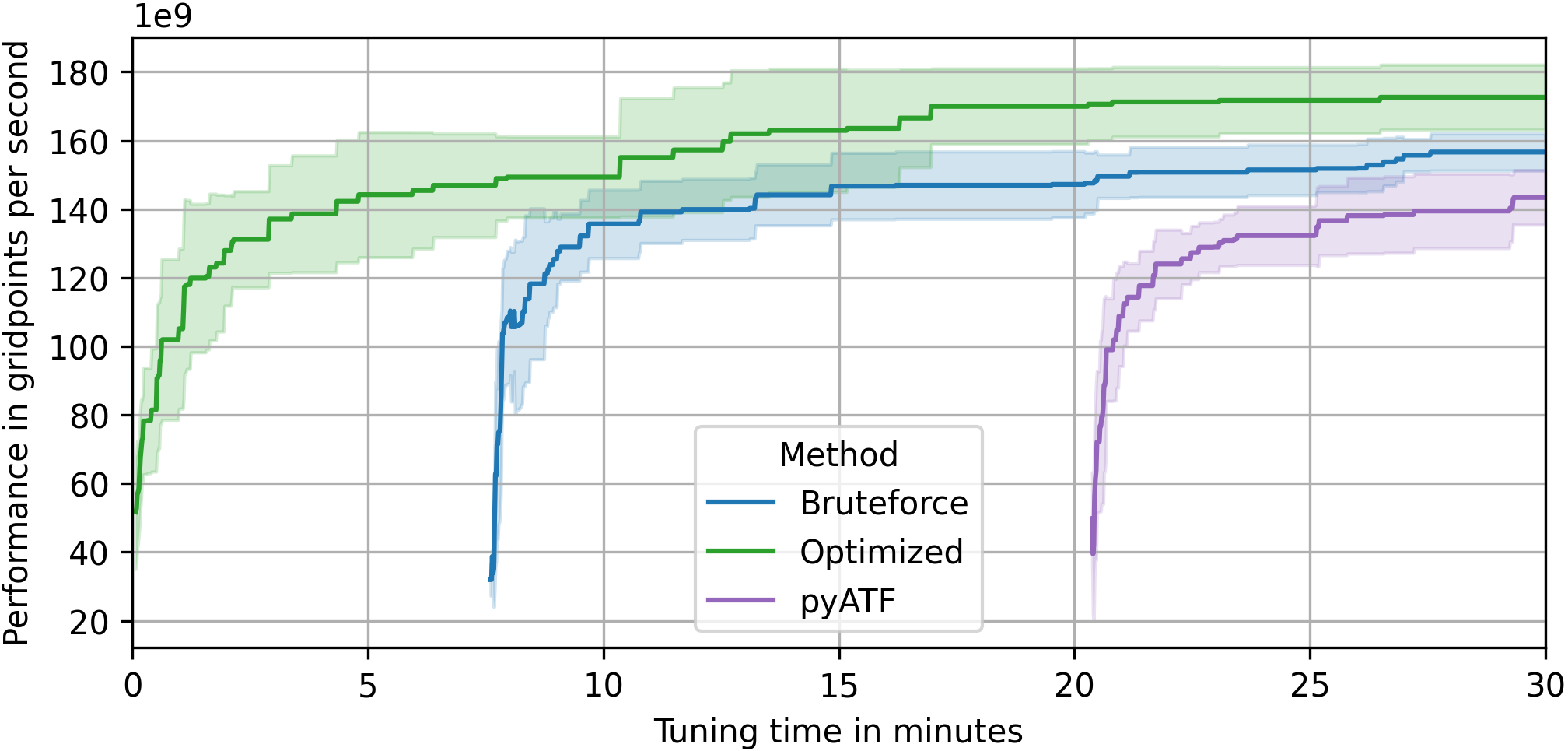}
    \vspace{-0.7cm}
    \caption{Best configuration performance found over a 30-minute auto-tuning of the \textit{hotspot} kernel using various search space construction methods.}
    \Description[Practical impact of search space construction method on auto-tuning runs]{Best configuration performance found over a 30-minute auto-tuning of the \textit{hotspot} kernel using various search space construction methods.}
    \label{fig:results_practical_impact_hotspot}
\end{figure}

\begin{figure}[tbp]
    \centering
    \vspace{-0.4cm}
    \includegraphics[width=1.0\linewidth]{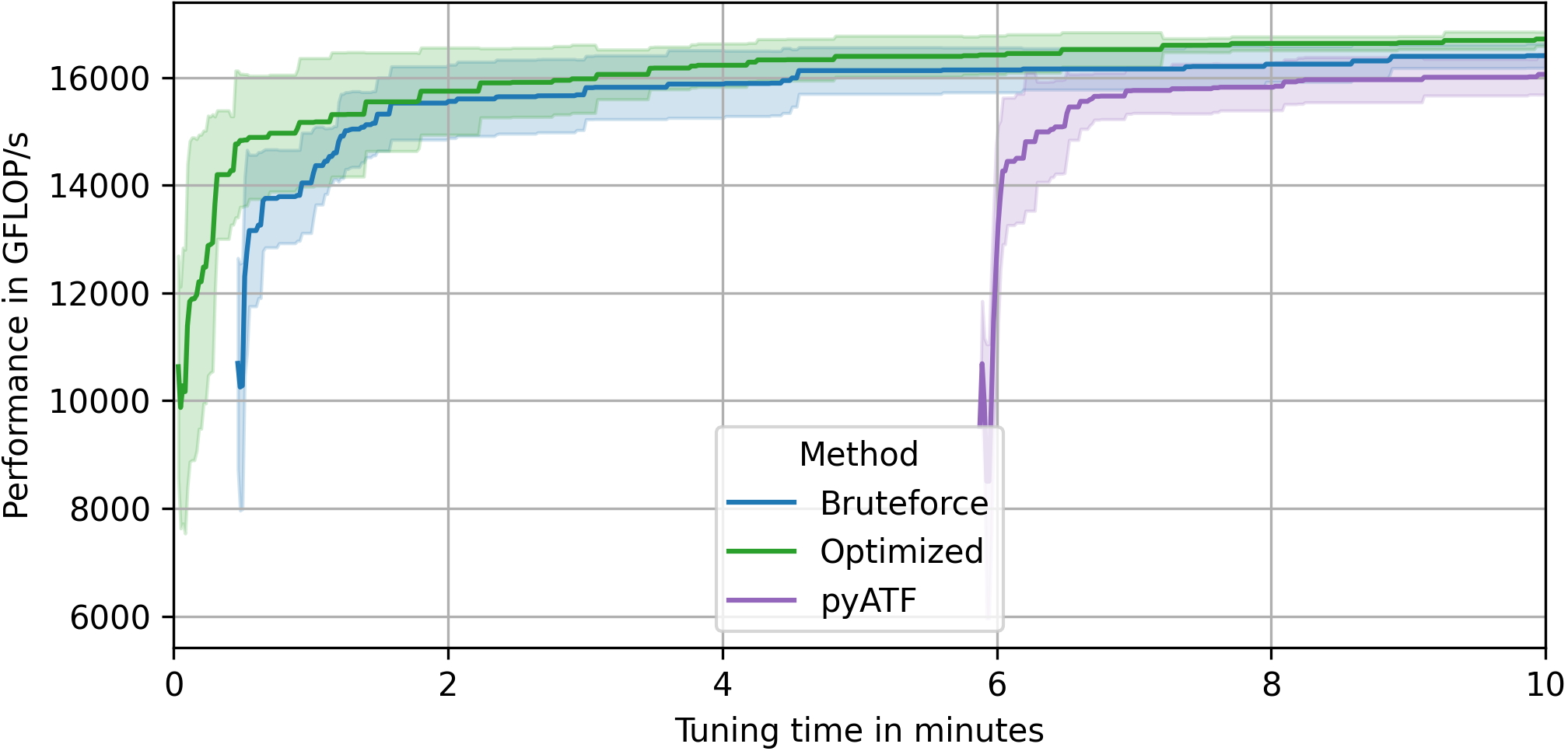}
    \vspace{-0.7cm}
    \caption{Best configuration performance found over a 10-minute auto-tuning of the \textit{GEMM} kernel using various search space construction methods.}
    \Description[Practical impact of search space construction method on auto-tuning runs]{Best configuration performance found over a 10-minute auto-tuning of the \textit{GEMM} kernel using various search space construction methods.}
    \label{fig:results_practical_impact_gemm}
\end{figure}

We auto-tune the \textit{hotspot} kernel described in \cref{subsubsec:evaluation_setup_kernel_hotspot} using the three Python-based solvers with a 30-minute time budget as an illustrative example. 
To avoid influence by a specific optimization algorithm, we use random sampling, and each run is repeated 10 times. 
\Cref{fig:results_practical_impact_hotspot} shows the best-performing configuration found so far during the tuning process, where higher is better. 
The time passed before any configuration is found is spent constructing the search space, which takes about eight minutes for brute-force and takes over twenty minutes for pyATF, while our optimized method is able to start to tune configurations almost immediately. 

To affirm these findings, we repeat this experiment on the \textit{GEMM} kernel, adjusting the time budget by the ratio between the number of valid configurations of GEMM and \textit{hotspot} seen in \cref{tab:searchspaces_real_world_overview} to 10 minutes. 
While brute force fares substantially better, as expected due to the smaller and denser search space, the results are otherwise very similar to those of the previous experiment. 
Most importantly, both examples confirm that the search space construction method has a substantial impact on the quality of the overall best configuration found within the time budget.

Overall, throughout this evaluation section, it is noteworthy that our optimized solver consistently outperforms any alternative on all of the search spaces by a wide margin, and the substantial practical impact this can have on the overall auto-tuning process. 
These findings emphasize the advantages of our optimized solver in efficiently handling large and complex search spaces.

    \section{Conclusions}
\label{sec:conclusion_futurework}

We introduced a novel approach to constructing auto-tuning search spaces using an optimized Constraint Satisfaction Problem (CSP) solver, addressing the specific challenges posed by the complexity of auto-tuning and large search spaces. 
Our contributions, available to the CSP-solving and auto-tuning community in the open-source \href{https://pypi.org/project/python-constraint2/}{python-constraint2} and \href{https://pypi.org/project/kernel-tuner/}{Kernel Tuner} packages, substantially outperform state-of-the-art methods in search space construction performance, enabling the exploration of previously unattainable problem scales in constraint-based auto-tuning and related domains.

Through rigorous evaluation, we demonstrated that our optimized CSP-based approach reduces construction time by several orders of magnitude, even for search spaces with billions of possible combinations. 
On average over the evaluated real-world applications, our optimized method is four orders of magnitude faster than brute force, three orders of magnitude faster than the unoptimized CSP solver, and one to two orders of magnitude faster than the state-of-the-art in search space construction. 
Our optimized search space construction method reduces the construction time of real-world applications to sub-second levels, eliminating it as a substantial factor in the overall tuning process overhead. %
In addition, our parsing method allows users to write constraints that are as close to the target language as possible, improving accessibility. 
Furthermore, our method prevents skewed sampling and has additional benefits for the efficiency of auto-tuning optimization algorithms. %

This breakthrough allows researchers and developers to more effectively harness the performance potential of modern hardware and provides an efficient generic solver for similar problem domains.

\textbf{Availability}: The methods presented in this work are available as user-friendly open-source software packages.
\iflackofspace
For more information, visit the \href{https://github.com/KernelTuner/kernel_tuner}{Kernel Tuner} and \href{https://github.com/python-constraint/python-constraint}{python-constraint} repositories. 
\fi

\ifdoubleblind
\else
\textbf{Acknowledgments}: The CORTEX project has received funding from the Dutch Research Council (NWO) in the framework of the NWA-ORC Call (file number NWA.1160.18.316).
\fi

\bibliographystyle{ACM-Reference-Format}
\bibliography{references}

\end{document}